# No evidence of superconductivity in the compressed sample prepared from the lutetium foil and $H_2/N_2$ gas mixture

Shu Cai[1,2]*, Jing Guo[1]*, Haiyun Shu[2], Liuxiang Yang[2], Pengyu Wang[1,3], Yazhou Zhou[1], Jinyu Zhao[1,3], Jinyu Han[1,3], Qi Wu[1], Wenge Yang[2], Tao Xiang[1,3], Ho-kwang Mao[2] and Liling Sun[1,2,3]†

[1]*Institute of Physics, Chinese Academy of Sciences, Beijing 100190, China*
[2]*Center for High Pressure Science & Technology Advanced Research, 100094 Beijing, China*
[3]*University of Chinese Academy of Sciences, Beijing 100190, China*

A material described as lutetium-hydrogen-nitrogen (Lu-H-N in short) was recently claimed to have "near-ambient superconductivity" [Gammon *et al*, Nature **615,** 244, 2023]. If the results could be reproduced by other teams, it would be a major scientific breakthrough. Here, we report our results of transport and structure measurements on a material prepared using the same method as that reported by Gammon *et al*. Our X-ray diffraction measurements indicated that the obtained sample contained three substances: the FCC-1 phase (Fm-3m) with a lattice parameter *a*=5.03 Å, the FCC-2 phase (Fm-3m) with a lattice parameter *a*= 4.755 Å and Lu metal. These two FCC phases are identical to the those reported in the so-called near-ambient superconductor. However, we found that the samples had no evidence of superconductivity, through our resistance measurements in the temperature range of 300 - 4 K and pressure range of 0.9 - 3.4 GPa, and our magnetic susceptibility measurements in the pressure range of 0.8-3.3 GPa and temperature down to 100 K. We also used a laser heating technique to heat the sample to 1800°C and found no superconductivity in the produced dark blue samples below 6.5 GPa. In addition, the color of the both samples remain dark blue in the pressure range investigated.

Room-temperature superconductors have been hunted by physicists for more than a century and are considered as a holy grail of condensed matter physics field. Since such material conducts electrons with zero resistance and expels magnetism at room temperature, it is expected to revolutionize our lives. Recently, Gammon *et al* reported evidence of near ambient superconductivity in a pink material called lutetium-hydrogen-nitrogen (Lu-H-N),[1] which attracts worldwide attention.[2-15] However, this finding arouses an intense debate because some related experiments performed recently[8,12] show that no superconductivity is observed in the sample synthesized by the alternative methods, such as employing the mixture of $NH_4Cl$ and $CaH_2$ as hydrogen and nitrogen sources and synthesizing the sample under high pressure and high temperature condition,[8,12] or taking ammonia borane as hydrogen and nitrogen sources and synthesizing the sample at low pressure with a laser heating technique.[3] In addition, the color change from blue to pink, considered a unique characteristic of this room temperature superconductor, has not been observed in their investigations, especially in the pressure range reported by Gammon *et al*.[1] In this study, we prepare the sample with the same procedure described in the Nature paper[1] and conduct our high-pressure resistance and magnetic susceptibility measurements on the produced sample, with the aim of clarifying the ongoing debate.

We loaded a lutetium foil (99.9%) with dimensions of 110 μm × 110 μm × 40 μm into the gasketed hole in a diamond anvil cell placed in a glovebox with argon atmosphere, and compressed the high pressure cell to ~2.3 GPa in the glovebox to isolate air and moisture. The high pressure cell was then taken out from the glovebox, and placed into a gas loading system. Prior to gas loading, the system was flushed three times with a mixture gas of compressed hydrogen (99%) and nitrogen (1%) to dilute the oxygen content in the chamber. The high-pressure cell with the preloaded Lu foil was opened in the gas loading system to allow gas mixture to enter into the sample hole. Then, the high-pressure cell was closed again in the gas loading system and pressurized to 2.0 GPa with a remote gear controlling device. Afterwards, the high pressure cell was heated to 65 °C in a furnace and kept at this temperature for 24 hours. The obtained sample appears dark-blue color under a transmitted white light (inset of Fig.1a), consistent with that reported by Gammon et al.[1]

Next, we performed high-pressure resistance measurements on the sample by the standard four-probe method. Figure 1a-1c show the temperature dependence of resistance in the pressure range covering the range reported by Gammon et al.[1] It was found that the resistance decreased upon cooling and exhibited a saturation behavior at lower temperatures, a typical metal behavior. Upon increasing pressure to 3.4 GPa, its resistance versus temperature exhibits the same trends. No drop in resistance was observed down to 4 K, indicating that the sample obtained at 65°C (we define this sample as "sample-obtained-at-65°C" in short) is not superconducting under these conditions. To determine its magnetic properties, we conducted high-pressure magnetic susceptibility measurements on the sample-obtained-at-65°C in a similar pressure range. As shown in Fig.1d-1f, temperature dependence of magnetization displays no diamagnetic signal, further confirming the absence of superconductivity in the studied sample.

We carried out X-ray diffraction (XRD) measurements on the sample-obtained-at-65°C, in order to check the chemical composition and crystal structure. As shown in Fig. 2, we found that the dark-bule sample is composed of three substances: one crystalizes in a face-centered-cubic (FCC) phase with space group Fm-3m and lattice parameter $a$=5.03 Å (defined it as FCC-1 phase, see the peak positions indicated by the red vertical bars), which corresponds to LuH$_2$; the second one hosts also a FCC unit cell (Fm-3m) with lattice parameter $a$= 4.755 Å (defined it as FCC-2 phase, see the peak positions indicated by the green vertical bars); the third one is determined as Lu hexagonal phase (P6$_3$/mmc) with lattice parameters $a$=3.516 Å, $b$= 3.516 Å, $c$=5.57 Å (see the peak positions indicated by the blue vertical bars). The first two phases are recognized as the same as the substances of "hydride compound A and B" described in the Nature paper.[1] Although Gammon *et al*[1] did not mention the third substance, Lu metal, in their XRD work, it is reasonable to have unreacted Lu metal remained in the sample because heating the sample to 65 °C and annealing the sample at this temperature for 24 hours cannot complete the reaction between the Lu foil and the gas mixture.

In order to determine the elemental composition of our sample, in particular nitrogen content, the energy dispersive spectroscopy (EDS) measurements were carried out. As shown in Fig.3, we measured eleven spots randomly on the sample-obtained-at-65°C (Fig.3a) and plotted the representative EDS spectra in Fig.3b. We did not find any trace of nitrogen is incorporated in our sample. The absence of nitrogen in the sample may be attributed to the small amount nitrogen (only 1%) in the gas mixture, which is too low to produce an effective reaction with the Lu foil. In fact, recent studies on the nitrogen- containing samples which are synthesized under high pressure and high temperature condition also find no superconductivity.[8,12] These results suggest that a small amount of nitrogen may be not crucial for developing the superconductivity in the Lu-H-N sample.

Given the fact that the heating temperature of 65 °C cannot make the lutetium foils to fully react with gas mixture (Fig.2), we thus employed a laser heating technique to heat another sample that is also prepared using the gas mixture of hydrogen and nitrogen sources to 1800 °C and performed the high-pressure resistance measurements. As shown in Fig.4a, the produced sample (we define this sample as "the sample-obtained-at-1800 °C" in short) displays a metallic behavior over the temperature range down to 4 K for pressures ranging from 0.9 GPa to 6.5 GPa. As a result, we conclude that the sample-obtained-at-1800 °C is not superconducting under these conditions, and the color of the compressed sample stays dark blue as well (Fig.4b and 4c). To check the crystal structure of the laser-heating sample, we conducted X-ray diffraction measurements on this sample. As shown in Fig.4d, the sample-obtained-at-1800 °C includes three substances: $LuH_2$ (see the peak positions indicated by the red vertical bars), FCC-2 phase (Fm-3m) with the lattice parameter $a$= 4.755 Å (see the peak positions indicated by the green vertical bars) and hexagonal $LuH_3$ ($P6_3/mmc$) with the lattice parameters $a$=3.56 Å, $b$= 3.56 Å, $c$=6.41 Å (see the peak positions indicated by the blue vertical bars). No Lu metal is found in the sample. These results indicate that the increase of the reaction temperature favors a thorough chemical reaction of the lutetium foil with the gas mixture.

$LuH_2$ and $LuH_3$ phases have been reported with no superconducting behavior in the temperature range of 300 K - 4 K under pressure up to 7.7 GPa for $LuH_2$ [2] and 122 GPa for $LuH_3$,[16] respectively, while the elemental Lu emerges a superconductivity at extremely low temperature (~0.3 K) and 11 GPa.[17] As for the FCC-2 phase found in this study is also not superconducting under pressure below 6.5 GPa.

Finally, an interesting phenomenon in the described room temperature superconductor is that the superconductivity occurs only in the pink phase, which transitions from the blue one in the pressure range of 0.3-3 GPa.[1] Although we employed the same method to produce the samples as described in the Nature paper, no color change is observed from our two compressed samples (the sample-obtained- at 65 °C and the sample-obtained-at 1800 °C) below 3.4 GPa and 6.5 GPa, respctively. The recent investigations showed that such a pink phase was observed in the non-superconducting Lu-H-N samples which were synthesized by high pressure and high temperature condition,[12,13] as well as in the pure $LuH_2$ sample in the pressure range of ~2 - ~5 GPa.[2,4] Such a pressure-induced color change has been known as piezochromism, which is associated with an alteration of the band gap, and often observed in plastics and semiconductors.[18,19] It may be an interesting topic for knowing whether such a material with piezochromism can host superconductivity.

In summary, we performed the high-pressure investigations on two samples initially prepared by the same method, using $99H_2/1N_2$ gas mixture and a lutetium foil as starting materials, followed by annealing the sample at 65°C for 24 hours as reported by Gammon *et al*[1] [Nature **615**, 247 (2023)], and heated the other one to 1800 °C for several minutes by the laser heating technique for comparison. The high-pressure resistance and magnetic susceptibility measurements on the sample-obtained-at-65°C did not show any evidence of superconductivity. The results of the ambient-pressure X-ray diffraction and energy dispersive spectroscopy experiments on this sample reveal that it contains three substances, including FCC-1 phase (Fm-3m) that corresponds to $LuH_2$, FCC-2 phase (Fm-3m) and hexagonal Lu metal ($P6_3$/mmc). These two FCC phases are identical to those found in so-called near-ambient superconductor described in the Nature paper [**615**, 247 (2023)]. No evidence of nitrogen incorporation in the sample-obtained-at-65°C is found. Our high-pressure resistance measurements on the sample-obtained-at-1800 °C (the first lutetium hydride synthesized by this method) also found no evidence of superconductivity when cooled down to 4 K. X-ray diffraction measurements showed that the sample-obtained-at-1800 °C consisted of hexagonal $LuH_3$ phase ($P6_3$/mmc), FCC-1 phase (Fm-3m) and FCC-2 phase. The color of both samples remained dark blue at pressure below 6.5 GPa. These studies allow us to propose that the samples, prepared by the same method as that reported in the Nature paper or synthesized by the laser heating method, do not exhibit superconductivity in the near-ambient temperature and pressure range described in the Nature paper.

These authors with star (*) contributed equally to this work.

Correspondence and requests for materials should be addressed to L.S. (llsun@iphy.ac.cn)

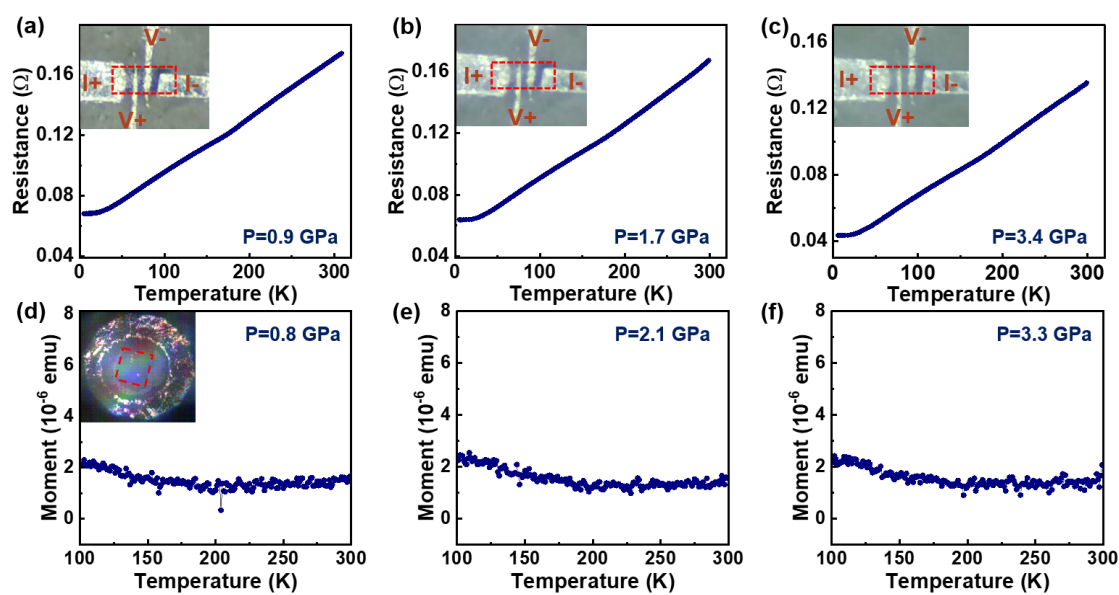

**Figure 1 The transport measurements for the sample-obtained-at-65 °C at low temperature and high pressure.** (a-c) Temperature dependence of resistance measured in the pressure range of 0.9 – 3.4 GPa. The insets show the arrangements of the sample and standard four probes, the sample is indicated by the red dash line. (d-f) Magnetic susceptibility versus temperature measured in the pressure range of 0.8-3.3 GPa and temperature rage of 100-300 K. The data are obtained with respect to the background subtraction. The inset of the figure (d) displays the sample in a gasket hole. Our resistance and magnetic results demonstrate that no evidence of superconductivity is found in the produced sample.

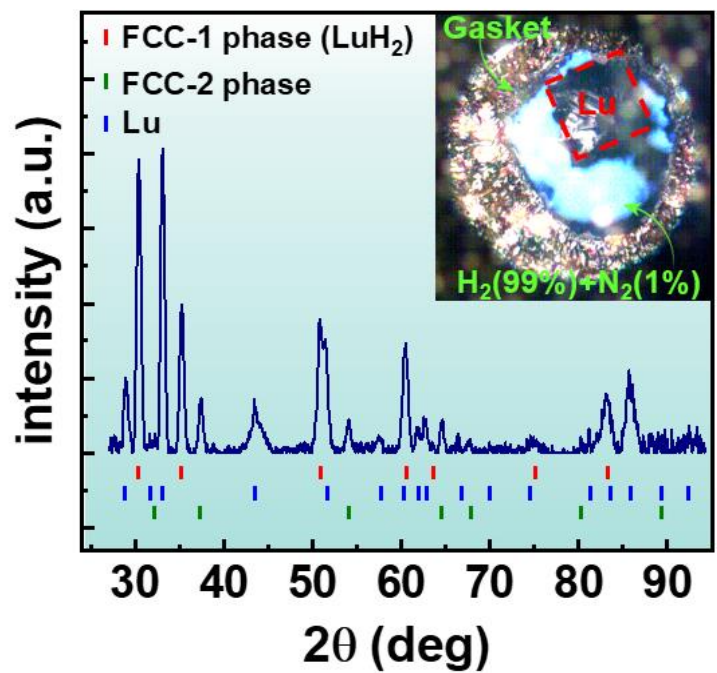

**Figure 2 X-ray diffraction analysis on the sample-obtained-at-65 °C.** FCC-1 phase

(Fm-3m) with lattice parameter $a$=5.03 Å (red bars), FCC-2 phase (Fm-3m) with lattice parameter $a$= 4.755 Å green bars) and Lu metal (blue bars) can be assigned to interpretate all diffraction peaks.

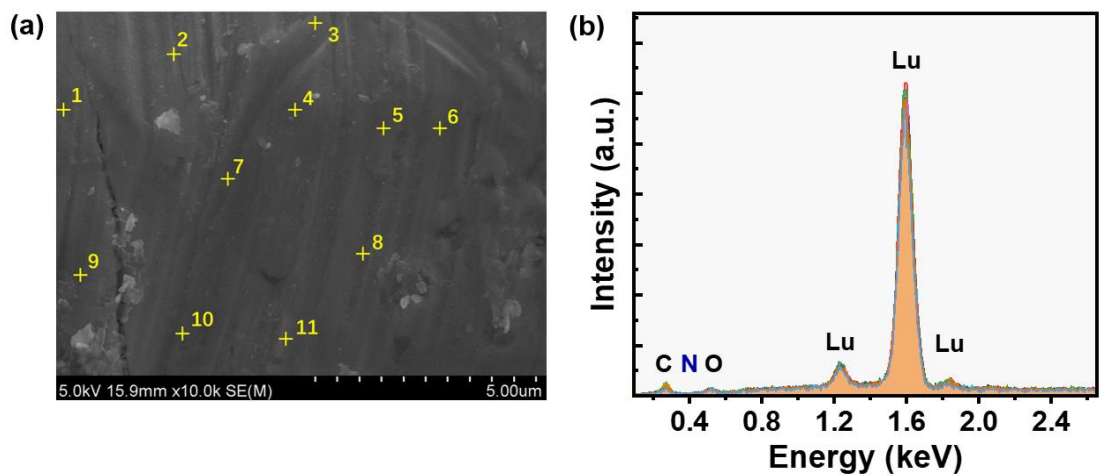

**Figure 3 Scanning electron microscopy (SEM) image and Energy dispersive spectrum (EDS) analysis on the sample-obtained-at-65 °C.** (a) SEM image taken from the produced sample surface. 11 locations (as marked in cross) were selected to collect the EDS. (b) The overlay of 11 EDS spectra measured shows no nitrogen is incorporated in the sample.

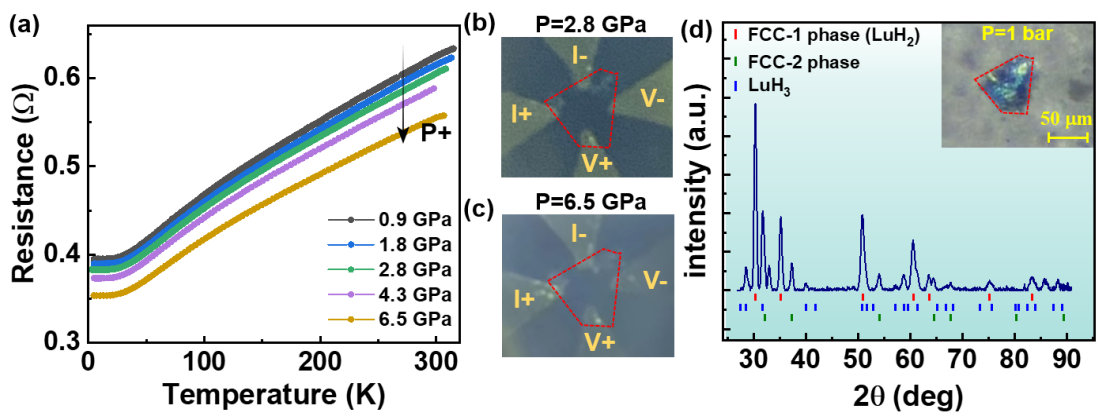

**Figure 4 The transport property and structural characterizations on the sample-obtained-at-1800 °C.** (a) Temperature dependence of resistance measured at different pressures, displaying no evidence of superconductivity in the experimental range of 0.9-6.5 GPa. (b) and (c) Photos taken at 2.8 GPa and 6.5 GPa, demonstrating that the sample's color remains dark blue. (c) The X-ray diffraction pattern from the produced sample, revealing the three substances, including $LuH_2$ (red bars), FCC-2 phase (Fm-3m, green bars) and $LuH_3$ (blue bars).